%% file: skeleton.tex
\documentclass{PoS}

\usepackage{amsmath}
\usepackage{amssymb}
\usepackage{subfigure}

\DeclareMathAlphabet{\mathcal}{OMS}{cmsy}{m}{n}

\graphicspath{{./}}

\input{newcommands.tex}

\title{Nucleon charges and quark momentum fraction with $N_f=2+1$ Wilson fermions}

\ShortTitle{Nucleon charges and quark momentum fraction with $N_f=2+1$ Wilson fermions}

\author{\speaker{Konstantin~Ottnad}$^{a}$, Tim~Harris$^{b,c,d}$, Harvey~Meyer$^{a,b}$, Georg~von~Hippel$^{a}$, Jonas~Wilhelm$^{a}$, Hartmut~Wittig$^{a,b}$ \\
        \llap{$^a$} PRISMA Cluster of Excellence and Institut f\"ur Kernphysik, Johann-Joachim-Becher-Weg 45, University of Mainz, 55099 Mainz, Germany \\
        \llap{$^b$} Helmholtz Institute Mainz, University of Mainz, 55099 Mainz, Germany \\
        \llap{$^c$} Dip. di Fisica G. Occhialini, Universita` degli Studi di Milano-Bicocca, Piazza della Scienza 3, 20126 Milano, Italy \\
        \llap{$^d$} INFN, Sezione di Milano-Bicocca Piazza della Scienza 3, I-20126 Milano, Italy \\\par
        E-mail: \email{kottnad@uni-mainz.de}, \email{tim.harris@mib.infn.it}, \email{meyerh@kph.uni-mainz.de}, \email{hippel@uni-mainz.de}, \email{jonas.wilhelm@uni-mainz.de}, \email{wittig@kph.uni-mainz.de}}

\abstract{We present a nucleon structure analysis including local isovector charges as well as twist-2 operator insertions. Computations are performed on CLS ensembles with $N_f=2+1$ Wilson fermions, covering four values of the lattice spacing and pion masses down to $M_\pi \approx 200\,\mev$. Several source-sink separations (typically $\sim 1.0\,\fm$ to $\sim 1.5\,\fm$) allow us to assess excited-state contaminations. Results on each ensemble are obtained from simultaneous two-state fits including all observables and all available source-sink separations with the mass gap as a free fit parameter. Finally, the chiral, continuum and finite size extrapolation is performed to extract physical results.}

\FullConference{The 36th Annual International Symposium on Lattice Field Theory - LATTICE2018\\
		22-28 July, 2018\\
		Michigan State University, East Lansing, Michigan, USA.}

\begin{document}

\section{Introduction}
We present a preliminary analysis of nucleon structure observables at zero momentum transfer performed by the Mainz group. For a previous account of this work see Ref.~\cite{Ottnad:2017mzd}. In this study we consider nucleon forward matrix elements 
\begin{equation}
 M_\mathcal{O}=\left<N(p',s') \left| \mathcal{O} \right| N(p,s)\right> \,,
 \label{eq:matrix_element}
\end{equation}
where $N(p,s)$, $N(p',s')$ denote the initial and final nucleon state and $\mathcal{O}$ is an (isovector) operator insertion. We use local operator insertions
\begin{equation}
 \mathcal{O}^A_\mu=\bar{q} \gamma_\mu \gamma_5 q, \quad \mathcal{O}^S=\bar{q}q , \quad \mathcal{O}^T_{\mu\nu} = \bar{q} i \sigma_{\mu\nu} q \,,
 \label{eq:local}
\end{equation}
giving rise to the isovector axial-, scalar- and tensor charges denoted by $g_A^{u-d}$, $g_S^{u-d}$, and $g_T^{u-d}$, respectively, as well as twist-2, dimension-four operator insertions
\begin{equation}
 \mathcal{O}^{vD}_{\mu\nu}  = \bar{q} \gamma_{\left\{ \mu \right.} \stackrel{\leftrightarrow}{D}_{\left.\nu\right\}} q \,, \quad \mathcal{O}^{aD}_{\mu\nu} = \bar{q} \gamma_{\left\{ \mu \right.} \gamma_5 \stackrel{\leftrightarrow}{D}_{\left.\nu\right\}} q \,, \quad \mathcal{O}^{tD}_{\mu\nu\rho} = \bar{q} \sigma_{\left[\mu\left\{\nu\right.\right]} \stackrel{\leftrightarrow}{D}_{\left.\rho\right\}} q \,.
 \label{eq:twist2}
\end{equation}
The later correspond to the nucleon isovector average quark momentum fraction  $\avgx{-}{}$, the helicity moment $\avgx{-}{\Delta}$ and the transversity moment $\avgx{-}{\delta}$, respectively. To compute physical observables within lattice QCD, we consider a ratio of spin-projected three- and two-point functions. For the relevant case of zero-momentum transfer $\vec{q}=\vec{p}'-\vec{p}=0$ it reads 
\begin{equation}
 R^\mathcal{O}_{\mu_1,...,\mu_n}(t_f,t,t_i)\equiv\frac{C_{\mu_1,...\mu_n}^{\mathcal{O},\mathrm{3pt}}(\vec{q}=0, t_f, t_i, t; \Gamma_z)}{C^\mathrm{2pt}(\vec{q}=0,t_f-t_i;\Gamma_0)} \,,
 \label{eq:ratio}
\end{equation}
where $t_i$, $t$ and $t_f$ denote initial state, insertion and final state times and the spin projectors $\Gamma_0=\frac{1}{2}(1+\gamma_0)$ and $\Gamma_z = \Gamma_0 (1+i\gamma_5\gamma_3)$ have been introduced. Considering large Euclidean time separations $t_f-t \gg 1$, $t-t_i\gg 1$ this ratio asymptotically reaches a plateau, which allows to extract the corresponding observable. However, in practice the signal-to-noise problem prevents one from choosing $\tsep=t_f-t_i \gg 1.5 \fm$, hence a different method is required which will be discussed below.

\section{Lattice setup}
Numerical calculations are performed on eleven gauge ensembles provided by the Coordinated Lattice Simulations (CLS) initiative. These ensembles have been generated with $N_f=2+1$ dynamical flavors of non-perturbatively improved Wilson quarks \cite{Bruno:2014jqa} and the tree-level Symanzik gauge action. Exceptional configurations are suppressed by a twisted mass regulator \cite{Luscher:2012av} and open boundary conditions in time are used to prevent issues with long autocorrelations in the topological charge~\cite{Luscher:2011kk}. An overview of the gauge ensembles can be found in Tab.~\ref{tab:ensembles}. In order to set the scale in our simulations we use the gluonic observable $t_0/a$ first introduced in Ref.~\cite{Luscher:2010iy} with a physical value of $\sqrt{8 t_0^\phys} = 0.415(4)_\stat(2)_\sys \fm$~\cite{Bruno:2016plf}. The ensembles comprise various choices of the light quark mass corresponding to pion masses between $\sim 200 \mev$ and $\sim 350 \mev$. For a reliable continuum extrapolation we have included simulations at four values of the lattice spacing $a$ between $0.0498\fm$ and $0.0863\fm$; see Ref.~\cite{Bruno:2016plf}. The ensembles used in this study typically satisfy $M_\pi L \gtrsim 4$, c.f. Tab.~\ref{tab:ensembles}. A single ensemble (S201) with $M_\pi L\approx 3$ has been added for a dedicated check of finite size effects. We remark that for two ensembles (C101, H102) we have only computed results for local operator insertions while for all other ensembles the full set of data is available. \par

The required renormalization of the bare matrix elements in Eq.~\ref{eq:matrix_element} has been performed non-perturbatively for the three coarser values of the lattice spacing using the Rome-Southampton method \cite{Martinelli:1994ty}. However, simulations with periodic boundary conditions as required by this method are not feasible at the finest lattice spacing due to the issue of topological freezing. Therefore, we use extrapolated values in this case.\par

For the computation of two- and three-point functions as required for the ratio in Eq.~(\ref{eq:ratio}) we employ the truncated solver method \cite{Bali:2009hu,Shintani:2014vja} on most of the ensembles. The corresponding numbers of high- and low-precision measurements $N_\mathrm{HP}$ and $N_\mathrm{LP}$ on each ensemble, as well as the available values of the source-sink separations $\tsep$ for the three-point functions are also included in Tab.~\ref{tab:ensembles}. Sequential inversions through the sink allow us to obtain the three-point function for all values of the insertion time $t$ for any given $\tsep$. The nucleon final state is always produced at rest, i.e. $\vec{p}'=0$. Since we restrict ourselves to isovector operator insertions, contributions from quark-disconnected diagrams cancel exactly. \par

Errors on observables are computed using the blocked jackknife method on each ensemble. The final errors from the chiral, continuum and finite size (CCF) fits are obtained via bootstrapping with $N=10000$ samples, after resampling the results on the individual ensembles before performing the fits. All errors on individual quantities (e.g. renormalization factors, $t_0/a^2$) are consistently propagated to the final results, hence we will quote only a single error reflecting both statistical and various systematic uncertainties. \par

\begin{table}[t]
 \setlength\tabcolsep{5.0pt}
 \centering
 \begin{tabular}{lrrrrrrrrl}
  \hline\hline
  ID  & $\beta$ & T/a & L/a & $aM_\pi$ & $M_\pi / \gev$ & $M_\pi L$ & $N_\mathrm{HP}$ & $N_\mathrm{LP}$ & $\tsep / \fm$ \\
  \hline\hline
  C101 & 3.40 &  96 & 48 & 0.0976(09) & 0.223(3) & 4.68 & 1908 & 15264 & 1.0, 1.2, 1.4           \\
  H102 & 3.40 &  96 & 32 & 0.1541(06) & 0.352(4) & 4.93 & 7988 &     0 & 1.0, 1.2, 1.4           \\
  H105 & 3.40 &  96 & 32 & 0.1219(10) & 0.278(4) & 3.90 & 4076 & 48912 & 1.0, 1.2, 1.4           \\
  \hline                                                  
  N401 & 3.46 & 128 & 48 & 0.1118(06) & 0.289(4) & 5.37 &  701 & 11216 & 1.1, 1.2, 1.4, 1.5, 1.7 \\
  S400 & 3.46 & 128 & 32 & 0.1352(06) & 0.350(4) & 4.33 & 1725 & 27600 & 1.1, 1.2, 1.4, 1.5, 1.7 \\
  \hline                                                  
  D200 & 3.55 & 128 & 64 & 0.0661(03) & 0.203(3) & 4.23 & 1021 & 32672 & 1.0, 1.2, 1.3, 1.4      \\
  N200 & 3.55 & 128 & 48 & 0.0920(03) & 0.283(3) & 4.42 & 1697 & 20364 & 1.0, 1.2, 1.3, 1.4      \\
  N203 & 3.55 & 128 & 48 & 0.1130(02) & 0.347(4) & 5.42 & 1540 & 24640 & 1.0, 1.2, 1.3, 1.4, 1.5 \\
  S201 & 3.55 & 128 & 32 & 0.0954(05) & 0.293(4) & 3.05 & 2092 & 66944 & 1.0, 1.2, 1.3, 1.4      \\
  \hline                                                  
  J303 & 3.70 & 192 & 64 & 0.0662(03) & 0.262(3) & 4.24 &  531 &  8496 & 1.0, 1.1, 1.2, 1.3      \\
  N302 & 3.70 & 128 & 48 & 0.0891(03) & 0.353(4) & 4.28 & 1177 & 18832 & 1.0, 1.1, 1.2, 1.3, 1.4 \\
  \hline\hline
  \vspace*{0.1cm}
 \end{tabular}
 \caption{List of CLS gauge ensembles used in this study with their respective values of $\beta$, the lattice spacing $a$, $T/a$ and $L/a$. The measured pion masses are given in units of the lattice spacing and in physical units. Additionally, we have included $M_\pi L$, the number of high and low precision measurements $N_\mathrm{HP}$, $N_\mathrm{LP}$ and the source-sink separations $\tsep$ in physical units.} 
 \label{tab:ensembles}
\end{table}

\section{Excited states and simultaneous fits}
\begin{figure}[t]
 \centering
 \includegraphics[totalheight=0.17\textheight]{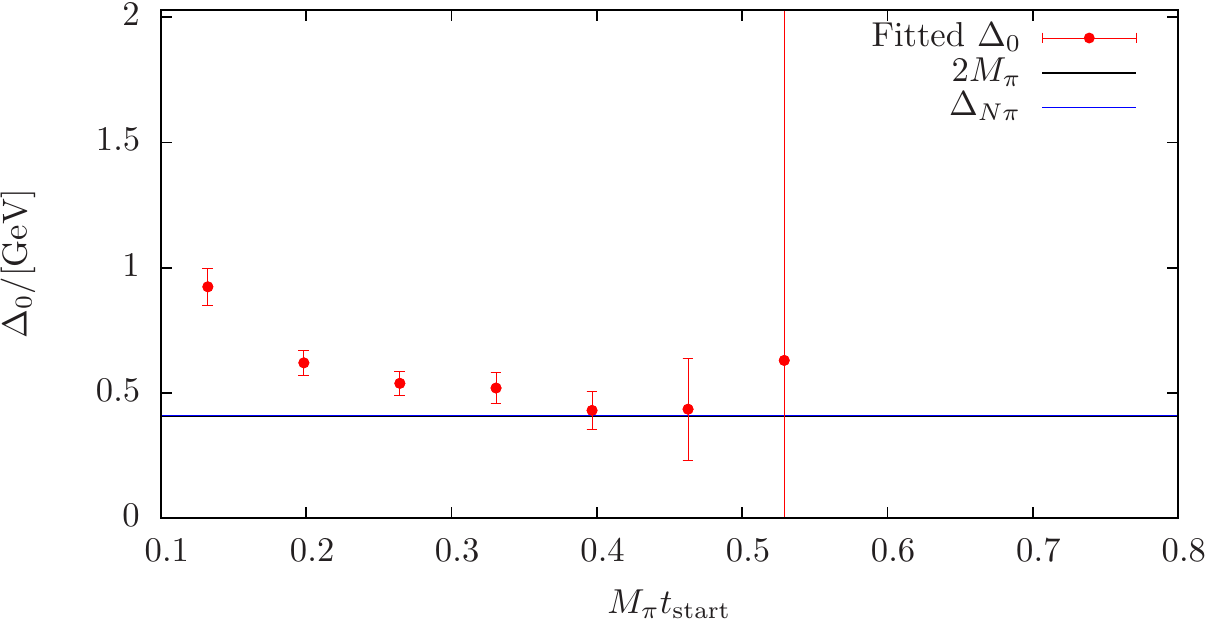}
 \includegraphics[totalheight=0.17\textheight]{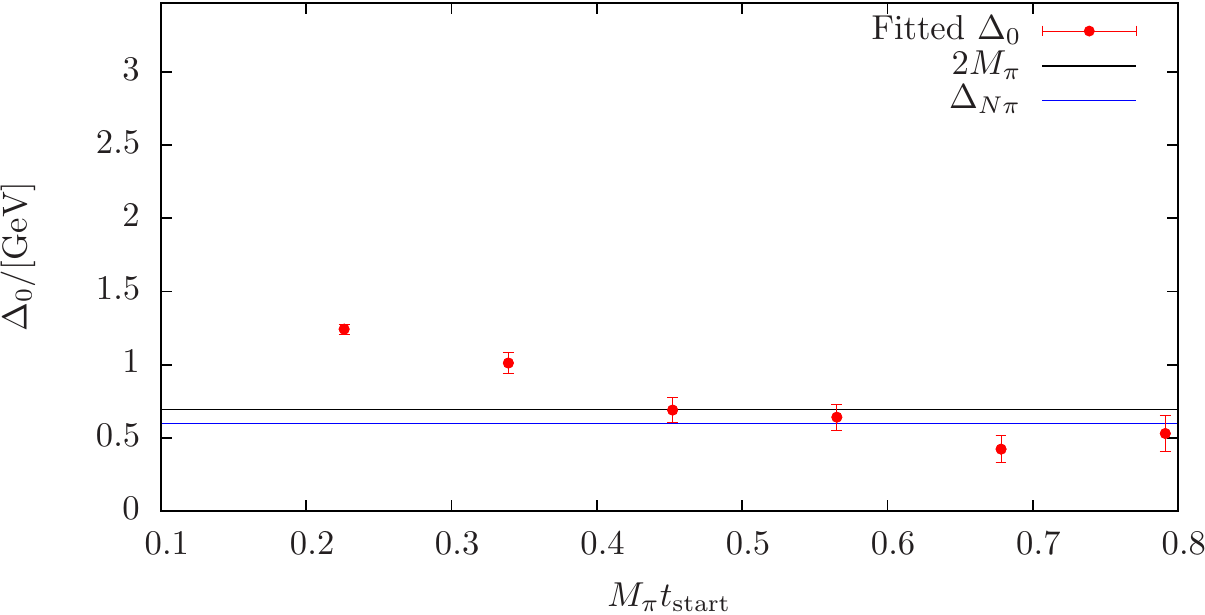}
 \caption{Convergence of the fitted gap $\Delta_0$ as a function of $M_\pi t_\mathrm{start}$. Left panel: Results on D200 without including the term $B_n^\mathcal{O} e^{-\Delta_n t_f}$  in Eq.~(\protect\ref{eq:fit_model}). Right panel: Results on N203 from the full fit model.}
 \label{fig:gap_convergence}
\end{figure}

\begin{figure}[thb]
 \centering
 \includegraphics[totalheight=0.17\textheight]{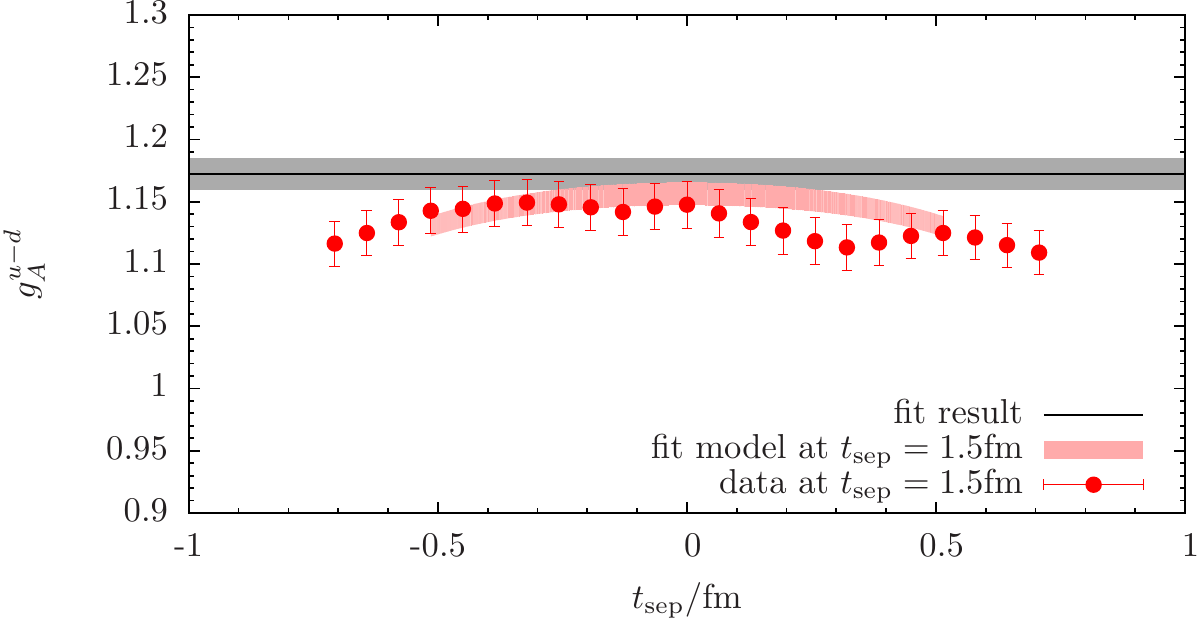}
 \includegraphics[totalheight=0.17\textheight]{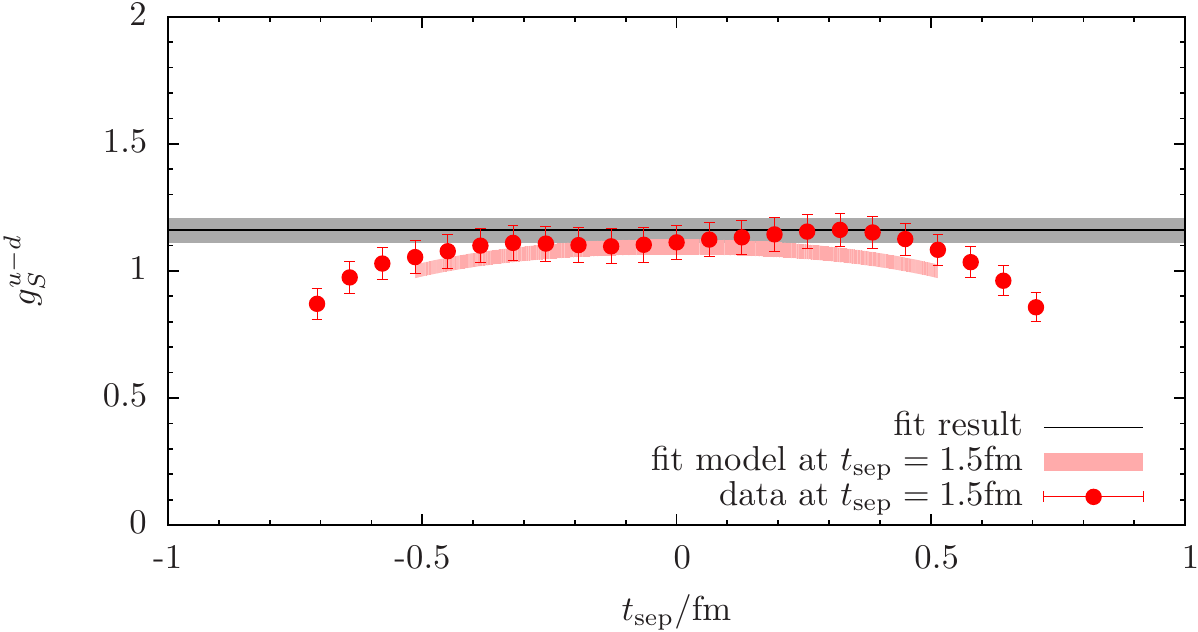} \\
 \includegraphics[totalheight=0.17\textheight]{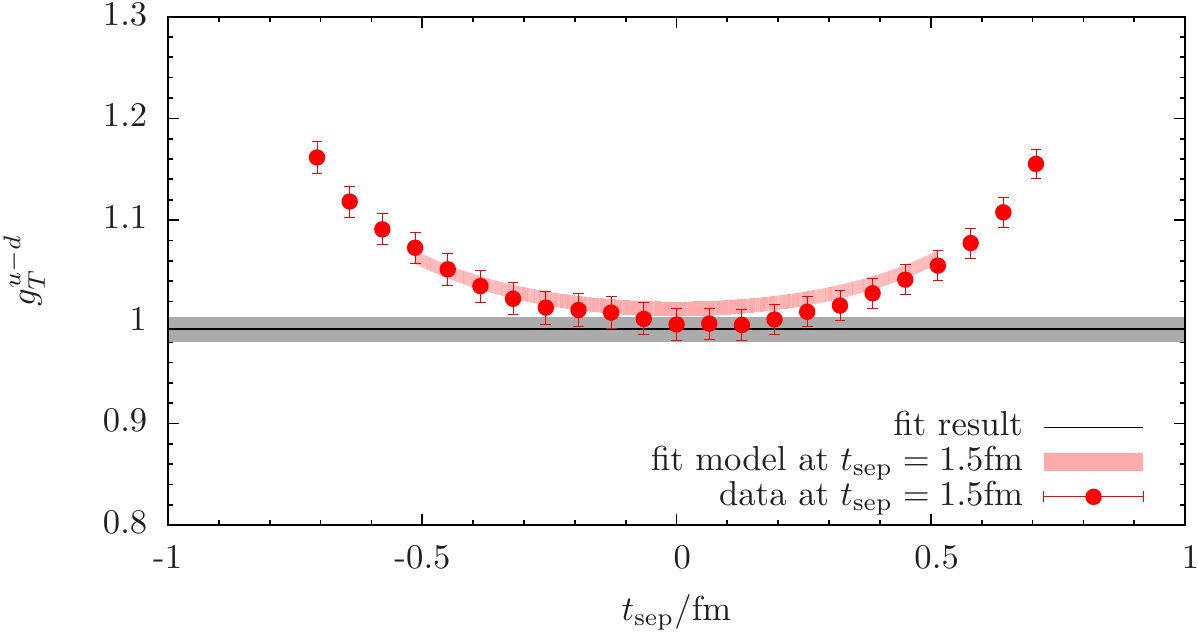}
 \includegraphics[totalheight=0.17\textheight]{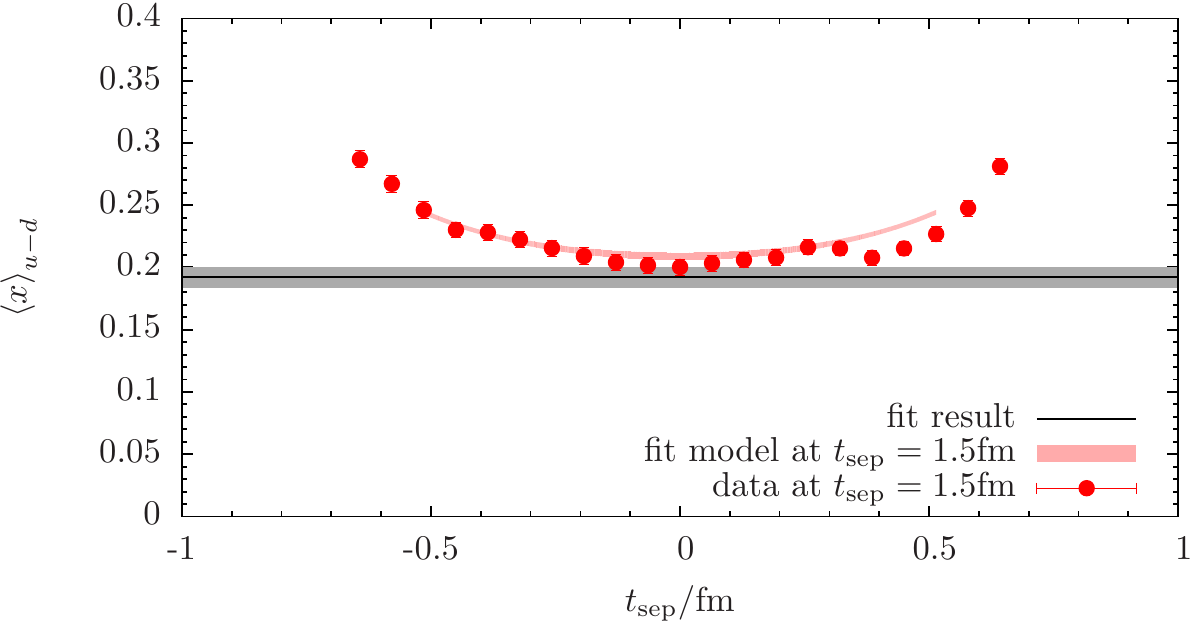} \\
 \includegraphics[totalheight=0.17\textheight]{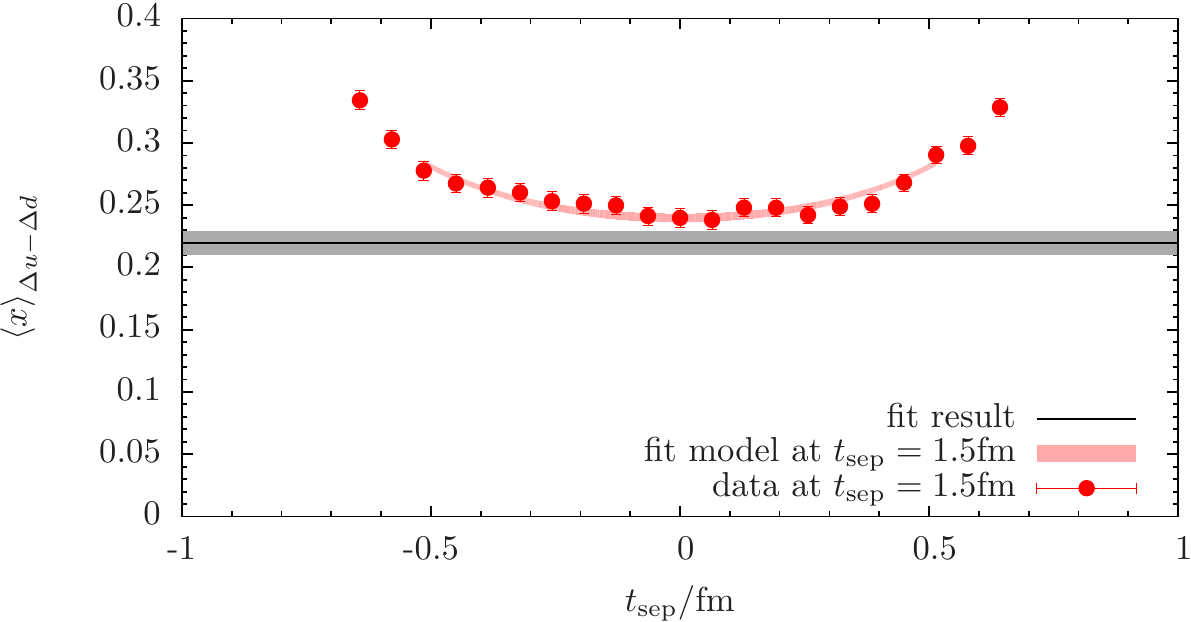}
 \includegraphics[totalheight=0.17\textheight]{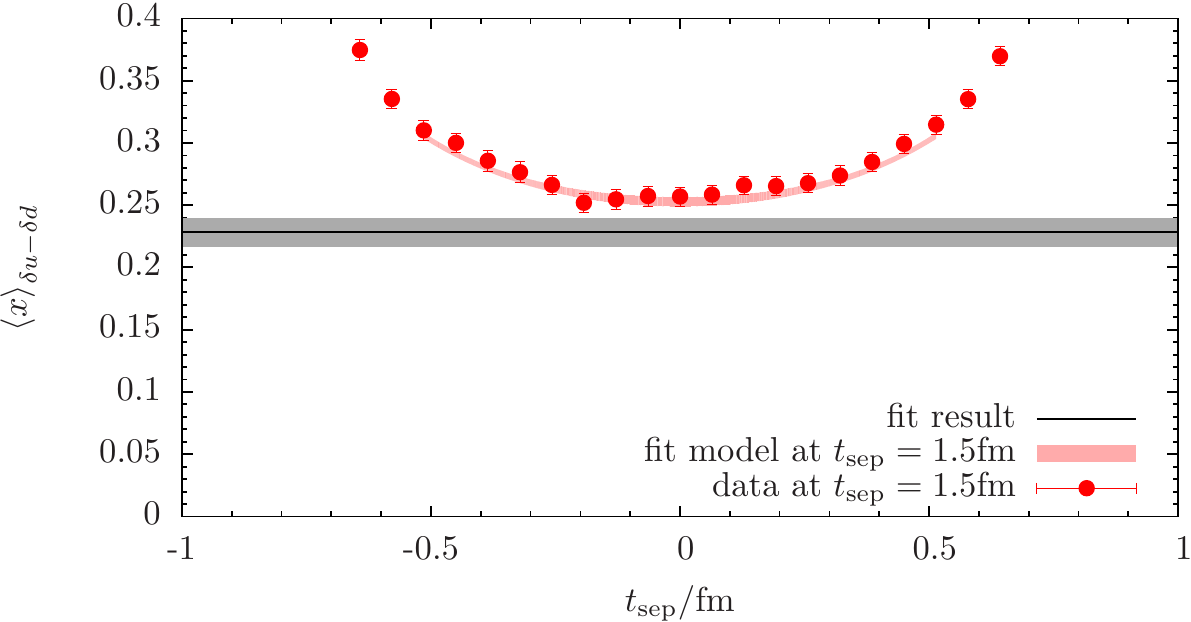}
 \caption{Results for all six observables on ensemble N203 from the simultaneous fit model in Eq.~(\protect\ref{eq:fit_model}). Lattice data and corresponding fit band (red) are only shown for the largest value of $\tsep\approx1.5\fm$.}
 \label{fig:simultaneous_fit}
\end{figure}

As mentioned before, lattice calculations of nucleon structure observables are strongly affected by excited state contamination, thus requiring a dedicated analysis to obtain physical results with reasonably controlled systematics. To this end we use a novel approach that allows us to apply a simultaneous fit directly to data for the ratio in Eq.~(\ref{eq:ratio}) including all available observables and all source-sink separations. Its validity relies on the fact that unlike amplitudes which differ for different operator insertions, the energy gaps are always the same. For the present case of zero-momentum transfer the fit ansatz reads
\begin{equation}
 R^{\mathcal{O}}(t_f,t,\vec{q}=0) = g_\mathcal{O} + \sum_n A_n^\mathcal{O} \left( e^{-\Delta_n t} + e^{-\Delta_n (t_f-t)}\right) + B_n^\mathcal{O} e^{-\Delta_n t_f} \,.
  \label{eq:fit_model}
  \end{equation}
where $g_\mathcal{O}$ is the actual charge (or moment), $\Delta_n$ is an energy gap and $A_n^\mathcal{O}$, $B_n^\mathcal{O}$ denote operator-dependent amplitudes. In our actual analysis we include a single state (i.e. one free gap $\Delta_0$) and perform a fit to symmetrized data in a range $\left[t_\mathrm{start} ,\tsep/2\right]$ for each value of $\tsep$ and all available observables. The value of $t_\mathrm{start}$ is chosen s.t. $M_\pi t_\mathrm{start}\approx0.4$ holds on each ensemble. In general, a precise result for the gap $\Delta_0$ at larger values of $t_\mathrm{start}$ requires very large statistics. Nevertheless, one can track the convergence of $\Delta_0$ as a function of $M_\pi t_\mathrm{start}$ on most of our ensembles. An example leaving out the last, suppressed term in the fit model is shown in the left panel of Fig.~\ref{fig:gap_convergence}. At sufficiently large statistics, this can be achieved even for the full model; see right panel of Fig.~\ref{fig:gap_convergence}. Typically, these fits describe the data rather well across all observables and values of $\tsep$, as can be seen in Fig.~\ref{fig:simultaneous_fit}. However, we might still have to adjust fit ranges in the future to better accommodate vastly different (effective) statistics on different ensembles, i.e. relax the strict criterion $M_\pi t_\mathrm{start}\approx0.4$. As can be inferred from comparing the actual fit results to the lattice data at the largest value of $\tsep$, there can still be significant effects of excited states even at source-sink separations as large as $\tsep=1.5\fm$ and even at fairly heavy pion masses ($M_\pi\approx347\mev$). \par

\section{Chiral, continuum and finite size extrapolations}
\begin{figure}[t]
 \centering
 \includegraphics[totalheight=0.19\textheight]{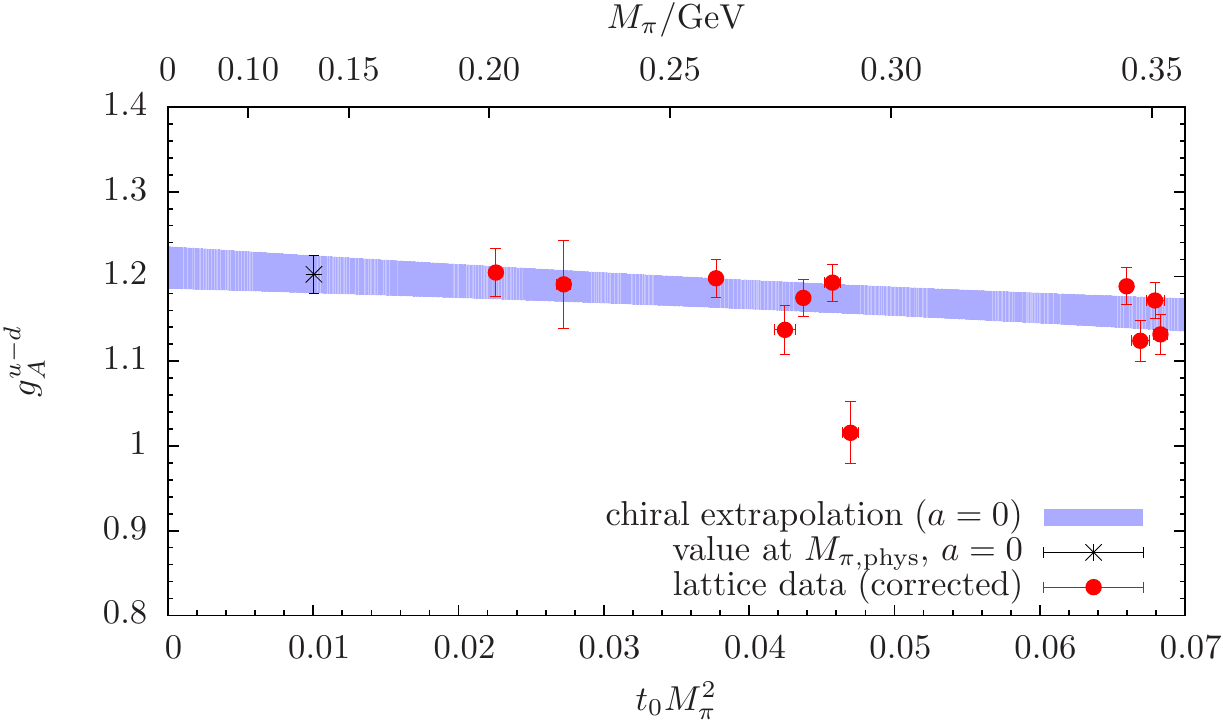}
 \includegraphics[totalheight=0.19\textheight]{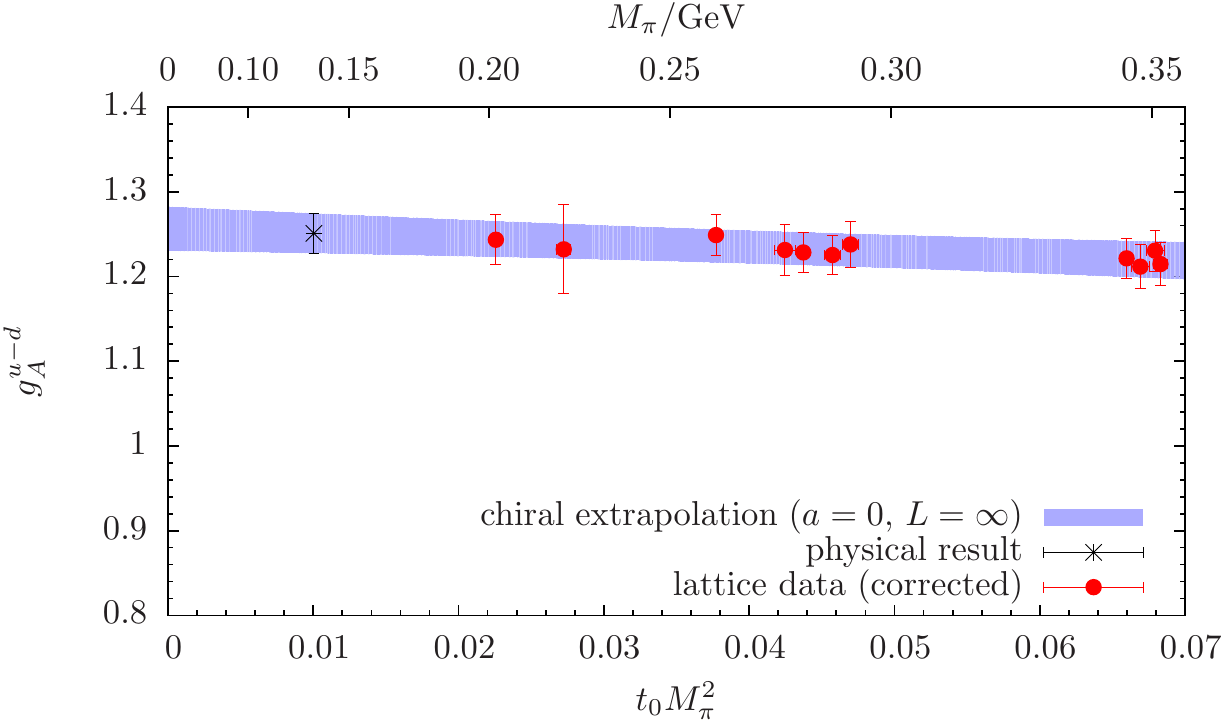} \\
 \includegraphics[totalheight=0.19\textheight]{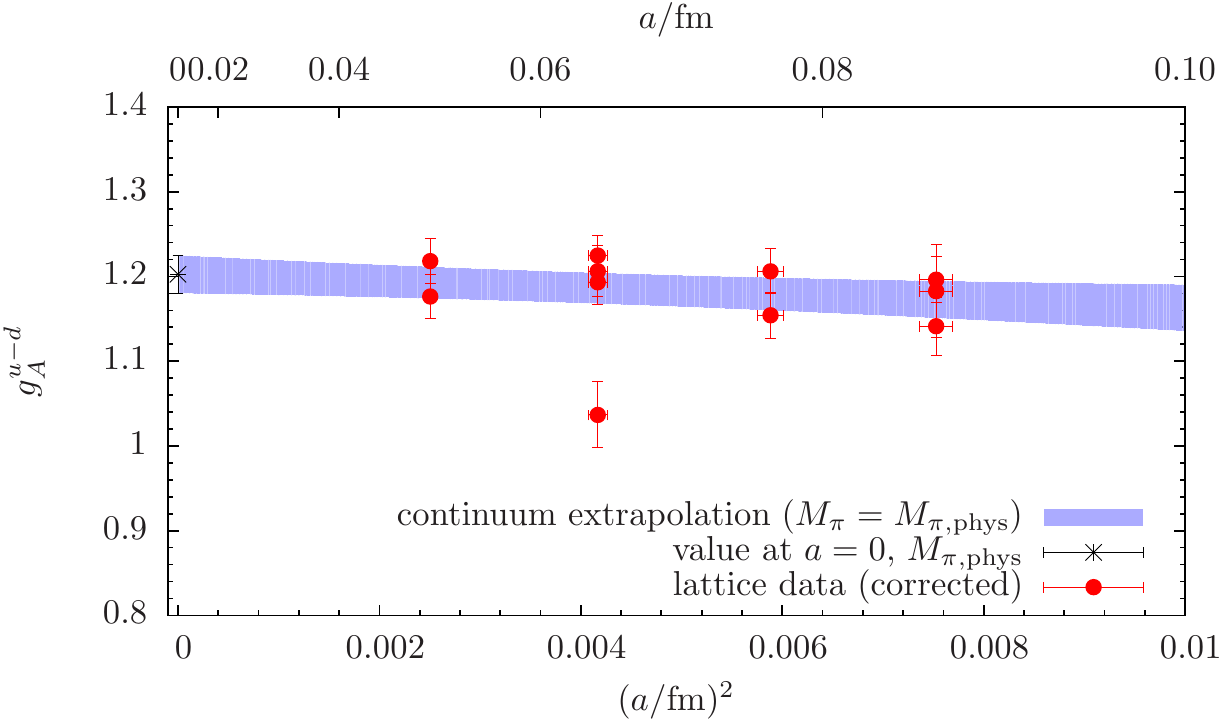}
 \includegraphics[totalheight=0.19\textheight]{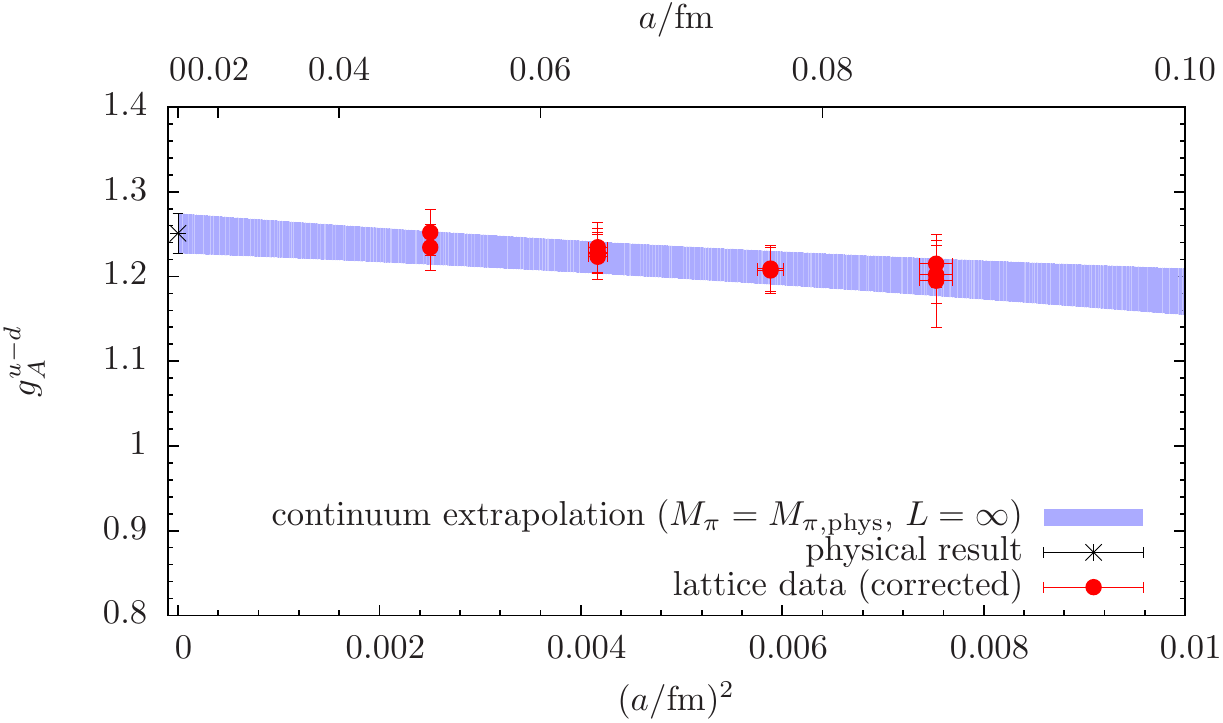}
 \caption{Chiral behavior (upper row) and continuum behavior (lower row) for $g_A^{u-d}$. Left column: Results from fit model ABD, i.e. no finite size term. Right column: Results from model ABDE.}
 \label{fig:g_A_CC_vs_CCF}
\end{figure}

\begin{figure}[thb]
 \centering
 \subfigure[Finite size behavior for $g_A^{u-d}$]{\includegraphics[totalheight=0.1655\textheight]{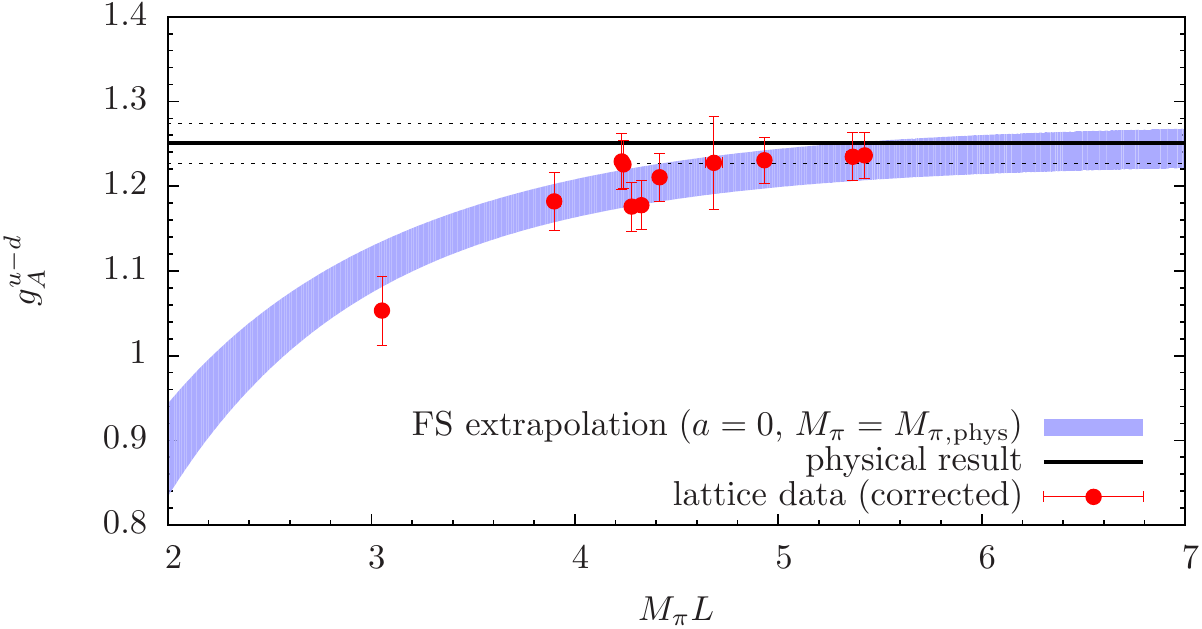}}
 \subfigure[Chiral behavior for $g_S^{u-d}$]{\includegraphics[totalheight=0.19\textheight]{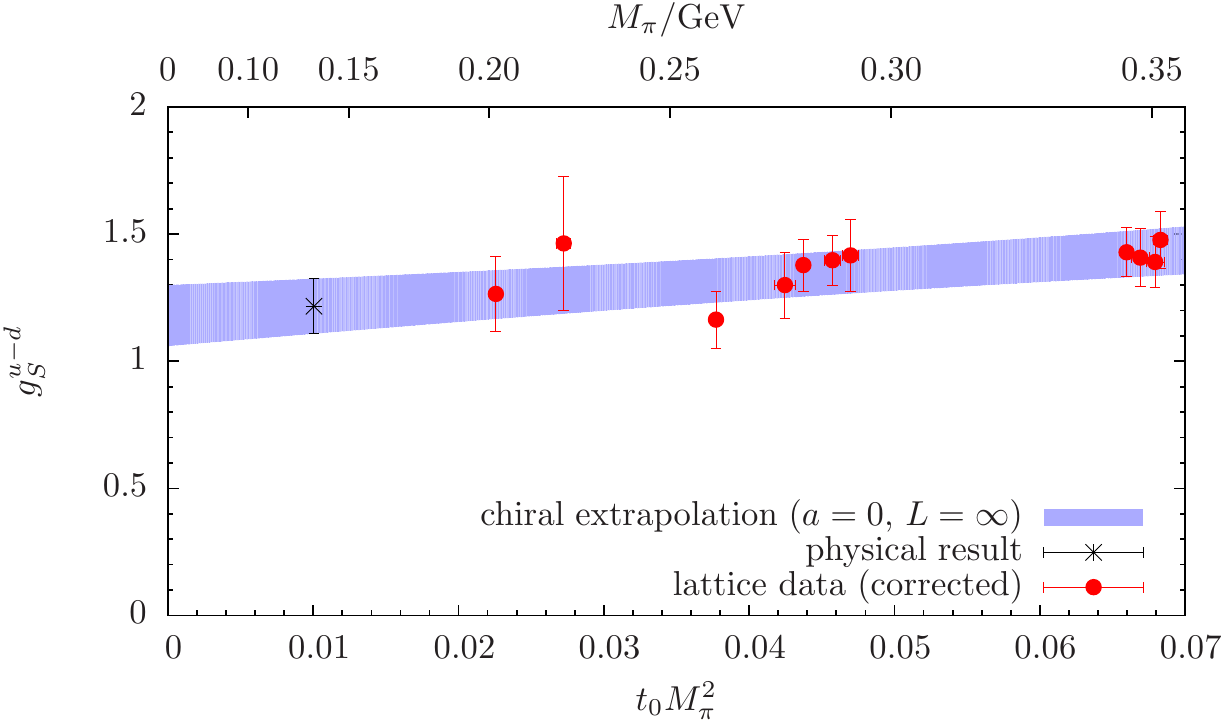}} \\
 \subfigure[Chiral behavior for $g_S^{u-d}$]{\includegraphics[totalheight=0.19\textheight]{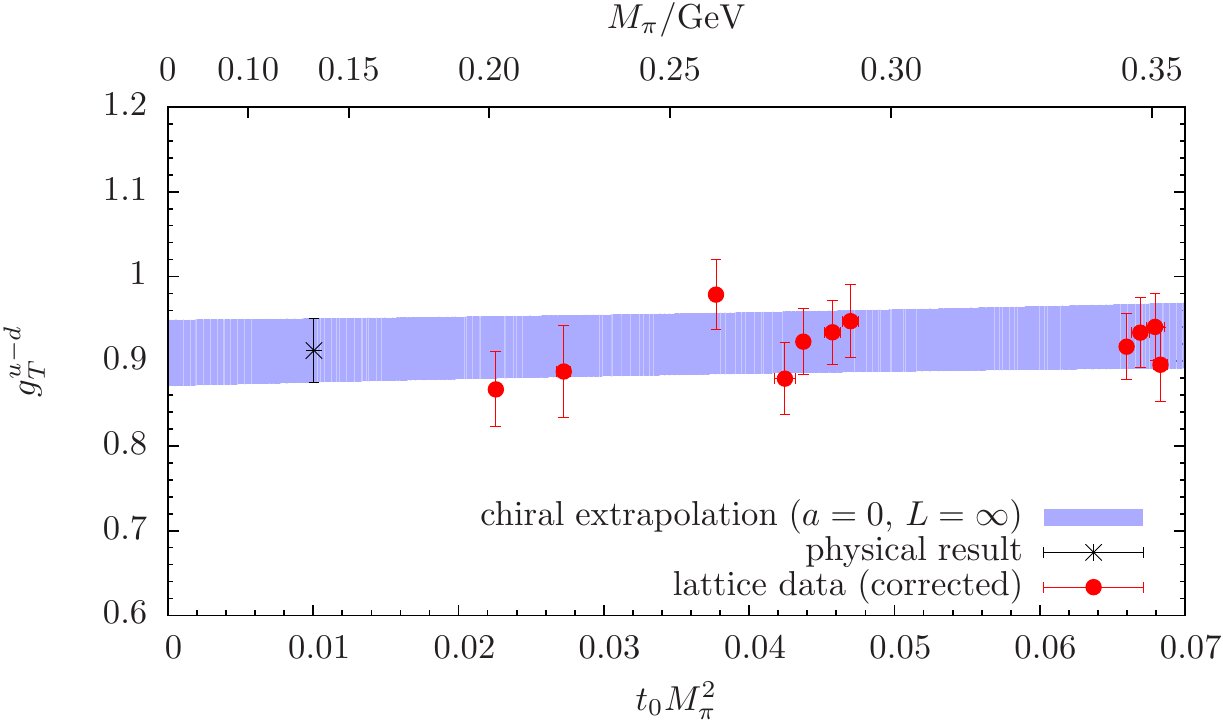}}
 \subfigure[Chiral behavior for $\avgx{-}{}$]{\includegraphics[totalheight=0.19\textheight]{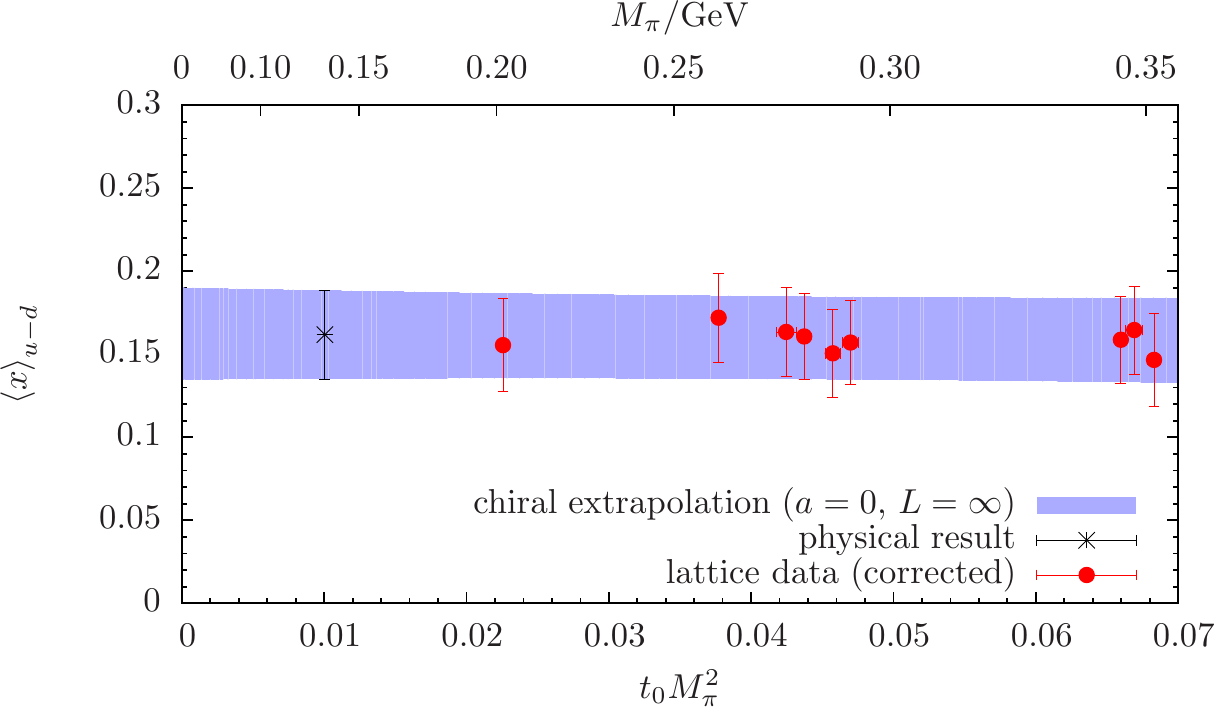}}
 \caption{Finite size and chiral behavior for several observables. All results from CCF fit model ABDE.}
 \label{fig:FS_and_CCF_extrapolation}
\end{figure}

In order to obtain physical results from the lattice data one still has to perform a chiral, continuum and finite size extrapolation. Therefore, we fit our data for each observable $O$ with the following fit model
\begin{equation}
  O(M_\pi,a,L) = A_O + B_O M_\pi^2 + C_O M_\pi^2 \log M_\pi + D_O a^{n(O)} + E_O \frac{M_\pi^2}{\sqrt{M_\pi L}} e^{-M_\pi L} \,,
 \label{eq:CCF_fit_model}
\end{equation}
where $n(O)$ controls the leading lattice artifact, i.e. $n(O)=2$ for $g_A^{u-d}$, $g_S^{u-d}$ and $n(O)=1$ otherwise. Individual fit models will be denoted by the corresponding combination of active fit parameters, e.g. 'ABDE'. In case of the axial charge the coefficient $C_O$ is known analytically, but we do not find our data to be sensitive to the chiral log in any case. This is why we do not include the term $\sim C_O$ in the final fits. Similarly, our present data does not constrain further higher-order terms. In Fig.~\ref{fig:g_A_CC_vs_CCF} we show the chiral and continuum behavior of $g_A^{u-d}$ for two choices of the fit model (ABD and ABDE). Note that in these plots the lattice data have been corrected for the remaining extrapolations using the parameters from the fits, leading to highly correlated point errors. We find that the inclusion of the finite size term generally improves the fit. In fact, for the axial charge finite size effects turn out to be quite significant, as can be seen from the finite size behavior for $g_A^{u-d}$ in the first panel of Fig.~\ref{fig:FS_and_CCF_extrapolation}. While mostly irrelevant for the final result, the additional S201 ensemble with $M_\pi L\approx 3$ clearly demonstrates the importance of the finite size correction. Therefore, we choose ABDE as our final fit model. The remaining panels of Fig.~\ref{fig:FS_and_CCF_extrapolation} show the chiral behavior for $g_S^{u-d}$, $g_T^{u-d}$ and $\avgx{-}{}$ (again after correcting the lattice data for all other effects). \par

\section{Results}
In general, we find that the data is described very well by the CCF fit model. Moreover, we have tested that applying a pion mass cut of $M_\pi<290\mev$ to our data does not significantly affect the results. The only exception is the tensor charge $g_T^{u-d}$ for which we obtain $\chi^2_\mathrm{corr} / \mathrm{dof}=2.53$ for fitting the full data set, but $\chi^2_\mathrm{corr} / \mathrm{dof}=1.50$ after applying the cut. Moreover, the resulting physical value changes by more than $1\sigma$. Therefore, we quote the more conservative value excluding data with $M_\pi>290\mev$ in this case. Our preliminary results for the isovector nucleon charges at the physical point read
\begin{equation}
 g_A^{u-d}=1.251(24) \,, \quad g_S^{u-d}=1.22(11) \,, \quad g_T^{u-d}=0.979(60) \,,
 \label{eq:final_results_charges}
\end{equation}
while for the lowest moments of the twist-2, dimension-four operators we find
\begin{equation}
 \avgx{-}{}=0.162(27) \,, \quad \avgx{-}{\Delta}=0.186(29) \,, \quad \avgx{-}{\delta}=0.169(38) \,.
 \label{eq:final_results_twist2}
\end{equation}

\section*{Acknowledgments}
This research is supported by the DFG through the SFB 1044. Calculations for this project were partly performed on the HPC clusters ``Clover'' at the Helmholtz-Institut Mainz and ``Mogon 2'' and  ``Himster 2'' at JGU Mainz. Additional computer time has been allocated through projects HMZ21 and HMZ36 on the BlueGene supercomputer system ``JUQUEEN'' at NIC, J\"ulich. We are grateful to our colleagues in the CLS initiative for sharing ensembles. We also would like to thank the Regensburg group for sharing additional ensembles that have been used in a collaborative effort for renormalization.

\end{document}

%% file: newcommands.tex
\newcommand{\gev}{\,\mathrm{GeV}}

\newcommand{\mev}{\,\mathrm{MeV}}
\newcommand{\fm}{\,\mathrm{fm}}

\newcommand{\phys}{\mathrm{phys}}
\newcommand{\stat}{\mathrm{stat}}
\newcommand{\sys}{\mathrm{sys}}

\newcommand{\tsep}{t_\mathrm{sep}}

\newcommand{\avgx}[2]{\langle x \rangle_{#2 u #1 #2 d}}